\documentclass[psfig]{mn2e}

\usepackage{graphicx}

\begin{document}

\title[A test of the failed disk wind scenario]
{A test of the failed disk wind scenario for the origin of the broad line region in active galactic nuclei}

\author[Galianni \& Horne]
{	Pasquale Galianni$^{1}$  and Keith Horne$^{1}$
\\	$^{1}$SUPA, School of Physics and Astronomy,
	The University of St.Andrews,
	North Haugh,
	St.Andrews KY16 9SS, Scotland, UK.
\\	(email: pg25,kdh1@st-andrews.ac.uk).
}

\date{Accepted . Received ; in original form }

\maketitle

\begin{abstract}

It has been recently proposed that the broad line region in active galactic nuclei originates from dusty clouds 
driven from the accretion disk by radiation pressure, at a distance from the black hole where the disk is cooler 
than the dust sublimation temperature. We test this scenario by checking the consistency of independent broad line 
region and accretion disk reverberation measurements, for a sample of 11 well studied active galactic nuclei. 
We show that {\it independent} disk and broad line region reverberation mapping measurements are compatible 
with a universal disk temperature at the H$\beta$ radius of $T(R_{\rm H\beta})\approx$ 1670 $\pm$ 231 K which is close 
to typical dust sublimation temperatures.
\end{abstract} 

\begin{keywords}
galaxies: active; quasars: general; Seyfert;
\end{keywords}

\section{Introduction}
\typeout{Intro}
\label{sec:intro}

Broad emission lines in the optical and UV bands are prominent features in many Active Galactic Nuclei (AGN) spectra. The Unified Model of AGNs \cite{ant85} assumes that broad emission lines seen in AGNs are generated in a zone called the Broad Line Region (BLR) located between a thick dusty torus and the central Black Hole (BH), in a region close to the accretion disk plane \cite{sug06,gas09}. Even though there are still some open problems regarding the physics and the geometry of the BLR (see Gaskell 2009 for a review) recent results seem to converge toward a scenario where the BLR is made of optically thick dusty clouds rotating around the BH in a roughly Keplerian fashion with the addition of some local turbulence (Done \& Krolik 1996, Bottorff et al. 1997, Marconi et al. 2008, Denney et al. 2009).
 Emission line reverberation mapping studies (Kaspi et al. 2000, Peterson et al. 2004, Bentz et al. 2009) show a tight relation between the H$\beta$ effective BLR radius and the monochromatic continuum luminosity at 5100 \AA :
 \begin{equation}
 R_{\rm H\beta}\propto \sqrt{L_{5100}}.
 \end{equation}
Although the $R_{\rm BLR}$-$L$ relation is well established and even used in an attempt to constrain cosmology (Watson et al. 2011) there is still an ongoing debate regarding its origin. Some of the proposed models for the BLR formation (Elvis 2000, Elitzur 2008) imply some kind of outflow from the disk that is either driven by radiation pressure (Chiang \& Murray 1996) or by the local magnetic field (Blandford \& Payne 1982).  \\
Recently, Czerny \& Hryniewicz (2011) (hereafter CH) proposed a simple formation mechanism for the BLR that naturally accounts for most of the observed properties of the BLR (e.g. the $R_{\rm BLR}-L$ relation, the observed turbulence in cloud motions and the universality of the $R_{\rm BLR}-L$ relation among narrow-line and non-narrow-line Seyferts 1). This "failed disk wind" scenario predicts that the BLR is made by dusty clouds driven above the accretion disk plane by radiation pressure, at radii where the underlying accretion disk is cooler than the dust sublimation temperature ($T_{\rm sub}\sim$ 1000 K). When the clouds rise at a sufficient height above the disk plane, the intense ionising radiation makes the dust sublimate. As a consequence, the clouds loose radiation pressure support and fall back on the disk plane. The mix of rising and falling clouds generates a kind of boiling motion.
This hypothesis effectively imposes a minimum radius to the inner edge of the BLR that is determined by the 
radius where the accretion disk has $T_{\rm eff}\approx T_{\rm sub}$.
In this paper we check the consistency between independent optical continuum (disk) and H$\beta$ (BLR) reverberation mapping measurements for a sample of 11 AGNs.
In Section 2 we describe the AGN sample and the dataset. In Section 3 we discuss our method. In Section 4 we show our results. Finally, we draw our conclusions in Section 5. 

\begin{table*}
 \centering
 \begin{minipage}{120mm}
  \caption{AGN data from disk and H$\beta$ reverberation mapping.}
  \begin{tabular}{@{}llccccc@{}}
  \hline
   Name & Type & \it{z} & $\log[M\dot{M}/{ M }_{ \odot  }^{ 2 }\rm yr^{-1}]$ \footnote{From Cackett, Horne \& Winkler (2007)}& $\tau(H_{\beta})[\rm days]$\footnote{From Bentz et al. (2009)} & $T(\rm R_{H\beta})$ [K]\\
    \hline
  NGC4051 & Sy 1 & 0.00234 & 5.06 $\pm$ 1.14 & 5.8$_{-1.8}^{+2.6}$ & 1183 $\pm$ 846\\ 
  \\
  NGC4151 & Sy 1.5 &0.00332 & 5.92 $\pm$ 0.16 & 6.6$_{-0.8}^{+1.1}$ & 1326 $\pm$ 187\\
  \\
  NGC3227 & Sy 2 & 0.00386 & 5.97 $\pm$ 0.40 & 7.8$_{-10.2}^{+3.5}$ &  2735 $\pm$ 1691\\
  \\
  NGC3516 & Sy 1.5 & 0.00884 & 6.96 $\pm$ 0.56 & 6.7$_{-3.8}^{+6.8}$ & 3281 $\pm$ 2142\\
  \\
  NGC7469 & Sy 1 & 0.01632 & 6.89 $\pm$ 0.36 & 4.5$_{-0.8}^{+0.7}$ & 3169 $\pm$ 769\\
  \\
  NGC5548 & Sy 1.5 & 0.01718 & 8.54 $\pm$ 0.23 & 18$_{-0.6}^{+0.6}$ & 2817 $\pm$ 384\\
  \\
  Mrk 79 & Sy 1 & 0.02219 & 7.49 $\pm$ 0.31 & 15.2$_{-5.1}^{+3.4}$ & 1868 $\pm$ 508\\
  \\
  Mrk 335 & Sy 1& 0.02579 & 7.60 $\pm$ 0.31 & 15.7$_{-4.}^{+3.4}$ & 1900 $\pm$ 479\\
  \\
  Ark 120 & Sy 1 & 0.03271 & 7.86 $\pm$ 0.34 & 39.7$_{-5.5}^{+3.9}$ & 1074 $\pm$ 231\\
  \\
  Mrk 509 & Sy 1.5 &0.0344 & 8.65 $\pm$ 0.31 & 79.6$_{-5.4}^{+6.1}$ & 998 $\pm$ 185\\
  \\
  3C390.3 & Sy 1.5 & 0.0561 & 7.65 $\pm$ 0.67 & 23.6$_{-6.7}^{+6.2}$ & 1543 $\pm$ 678\\  
\hline
\end{tabular}
\end{minipage}
\end{table*}

\section{The AGN sample}

Our sample consists of 11 bright AGN for which there are both disk (Cackett, Horne \& Winkler 2007) and H$\beta$ (Bentz et. al 2009) reverberation mapping measurements (see Table 1).
Our sample consists mostly of Type 1 Seyfert galaxies, and spans a range of redshift between $z\sim$0.002 and $z\sim$0.06.
For the disk reverberation we use results from Cackett, Horne \& Winkler (2007) where the authors measured
time delays for a sample of 14 AGN observed in optical BVRI bands by Sergeev et al. (2005). The authors fit a reddened thin accretion disk model (Shakura \& Sunyaev 1973)  to the variable component of their light and obtain optical continuum time delays by cross-correlation analysis of the light curves. For the H$\beta$ reverberation we use results from Bentz et. al (2009) where host galaxy contamination was carefully taken into account using high resolution HST images.

\section{Method}

We assume a simple thin steady-state accretion disk model (Shakura \& Sunyaev 1973) 

\begin{equation}
R=\left( \frac { 3GM\dot { M } }{ 8\pi \sigma _{ s } T^{ 4 } }  \right)^{1/3}, 
\end{equation}
where $T$ and $R$ are the temperature and radius respectively, $M$ the BH mass, $\dot{M}$ the accretion 
rate and $\sigma _{ s }$ the Stefan-Boltzmann constant. \\
The scenario proposed by CH predicts a disk temperature at the H$\beta$ radius that is
universal among AGN and close to typical dust sublimation temperatures (1000 K $<T_{\rm sub}$ $<$ 2000 K):

\begin{equation}
T_{\rm eff }(R_{H\beta}) \approx T_{\rm sub}.
\end{equation}

Parametrized in terms of the H${\beta}$ time lag $\tau=R_{H\beta}/c$ and the $M\dot{M}$ product, 
Equations (2) and (3) translate to the following relation

\begin{equation}
\log(\tau) \approx 3.001  - \frac{4}{3} \log(T) + \frac{1}{3} \log(M\dot{M}) ,
\end{equation}
where $\tau$ is in days and $M\dot{M}$ in $M_\odot^2 \rm yr^{-1}$. We use the above relation to test the predictions of the failed disk wind scenario and the consistency of disk/BLR reverberation measurements.

\begin{figure}
\includegraphics[width=7cm,angle=270]{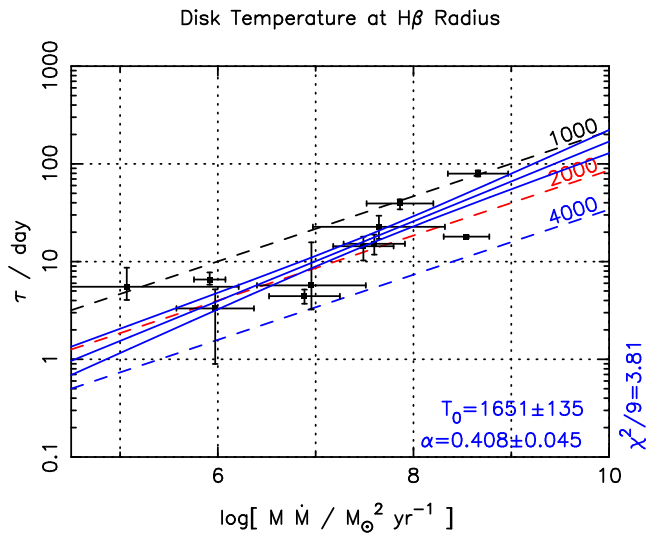}
\caption{H$\beta$ time delays plotted against accretion rates from Table 1. The continuous lines represent the best fit 2-parameter model with a 1-$\sigma$ error envelope. 
Dashed lines represent Eq. (4)  plotted for 3 different temperatures.
Here the slope of the linear relation $\alpha$ is determined as a free parameter and looks compatible with the theoretical slope of 1/3 (see text for details).}
\label{fig1}
\end{figure}

\begin{figure}
\includegraphics[width=7.5cm,angle=0]{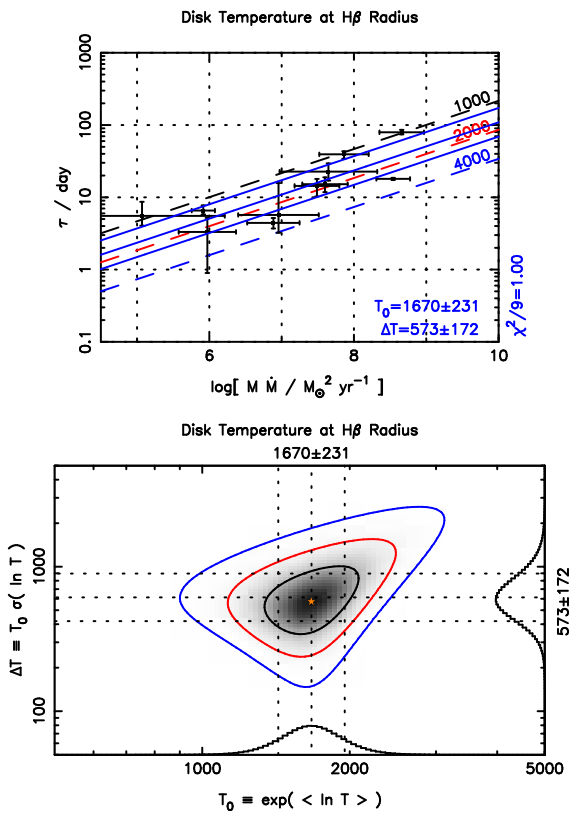}
\caption{ (upper panel) 2-parameter maximum likelihood fit to the data in Table 1. Here the two free parameters are a centroid temperature $T_0$ and a temperature dispersion $\Delta T$. The continuous lines represent the best fit model and the span due to the intrinsic scatter. The dashed lines show Eq. 4 plotted for 3 different temperatures. (lower panel) posterior probability distribution of the parameters. The contours represent 1, 2 and 3-$\sigma$ confidence regions.}
\label{fig2}
\end{figure}

\section{Data analysis and results}

 Our first fit considers a power-law model for the time delay $\tau$ as a function of $M\dot{M}$. Fig.~1
shows the data points and error bars from Table~1 transformed onto logarithmic scales.
When asymmetric error bars are given for $\tau$, we use their geometric mean. A correlation is evident
in Fig.~1, and the slope can be judged by eye to be close to the theoretical slope of $1/3$,
as indicated by the dashed curves for three disk temperatures. 
In Fig 1 we fit a 2-parameter linear relationship between $\log(\tau)$ and
$\log(M\dot{M})$, accounting for the error bars on both axes.
As shown in Fig~1, we find a best-fit slope of $\alpha=0.408\pm0.045$.
As this is only $1.6~\sigma$ steeper than the theoretical slope $1/3$,
we conclude that the data are consistent with the theory.
The reduced $\chi^2$ (3.81 with 9 degrees of freedom) indicates a larger scatter
among the data points than can be accounted for
by the observational errors. Thus if the error bars are reliable,
our fits provide some evidence that the disk temperature
at the H$\beta$ radius varies from object to object. 
To quantify the object-to-object distribution of disk
temperatures at the H$\beta$ radius, we define a 2-parameter model with centroid temperature
        $T_0 \equiv \exp{( \left< \ln{T} \right> )}$
and a temperature dispersion
        $\Delta T \equiv T_0~\sigma\left( \ln{T} \right)$.
Our maximum likelihood fit, shown in Fig.~2, gives
$T_0=1670\pm231$ and $\Delta T=573\pm172$.
The reduced $\chi^2/9$ is now 1, compatible with the scatter
by virtue of the intrinsic dispersion.
The parameters are well enough defined that the posterior
distribution is gaussian (lower panel of Fig.2).
Our results indicate that the accretion disk temperature
at the H$\beta$ radius is $T(R_{\rm H\beta})\approx 1670\pm231$ K which is 
quite close to typical dust sublimation temperatures.

This value is $\approx$ 700 K higher than that obtained
by CH (994 $\pm$ 74 K). This difference can be largely accounted for by the differences
in the assumed H${\beta}$ reprocessing geometry between our and CH work.
The time delay at radius R is
\begin{equation}
\tau = (R/c) ( 1 + \sin{i} \cos{\theta} ),
\end{equation}

where $i$ is the inclination to the line of sight and $\theta$ the azimuthal angle.
CH assumes a mean BLR inclination of $i$=39.2$^{\circ}$ and that the H${\beta}$ emission arises primarly from the 
part of the BLR farthest away from the observer such that $\cos{\theta}\approx$ 1. With these assumptions the time 
delay increases by a factor $\tau = 1.63~(R/c)$. We assume instead that the H${\beta}$ response is roughly uniformly 
distributed in azimuth such that $\langle\cos{\theta}\rangle$= 0. Considering the $T \propto R^{-3/4}$ relation, 
one finds that our temperature is $(1+\sin{i})^{0.75}$= 1.44 times higher than in the case of emission coming from 
the farthest side of the BLR. Correcting by this factor yield a temperature centroid of 
$T(R_{\rm H\beta})\approx 1160$ K which is much closer to CH value.

The intrinsic temperature scatter of $\Delta T=573\pm172$ is not surprising, considering 
that H$\beta$ delays and $M\dot{M}$ measurements are not simultaneous and thus do not correspond
to the same disk luminosity.

\section{Conclusions}

We have shown that {\it independent} disk and H$\beta$ reverberation mapping measurements are compatible 
with a universal disk temperature at the H$\beta$ radius of $T(R_{\rm H\beta})\approx$ 1670 $\pm$ 231 K which is close 
to typical dust sublimation temperatures.
Our results are therefore compatible with the failed disk wind scenario for the origin of the BLR proposed by Czerny \& Hryniewicz (2011). In this framework the $R_{\rm BLR}$-$L$ scaling relation can be interpreted as  a consequence of the accretion disk luminosity-radius scaling explored in Cackett, Horne \& Winkler (2007) and the existence of a universal dust sublimation temperature among AGN.

\subsection*{Acknowledgements}
KH is supported by a Royal Society Leverhulme Trust Research fellowship. PG is supported by an STFC studentship.



\begin{thebibliography}{}

  \bibitem[\protect\citename{Antonucci \& Miller} 1985]{ant85}
    Antonucci R. R. J., Miller J. S.,
    1995, ApJ, 297, 621.
  
  \bibitem[\protect\citename{Bentz et al.} 2009]{ben09}
    Bentz M. C. et al.,
    2009, ApJ, 697, 160.
     
  \bibitem[\protect\citename{Bottorff et al.} 1997]{bot97}
    Bottorff M. et al.,
    1997, ApJ, 479, 200.
  
  \bibitem[\protect\citename{Blandford \& Payne} 1982]{blan82}
    Blanford R. D., Payne, D. G.
    1982, MNRAS, 199, 883.
     
  \bibitem[\protect\citename{Cackett, Horne \& Winkler} 2007]{cac07}
   Cackett E. M., Horne K., Winkler H.,
   2007, MNRAS, 380, 669.
   
  \bibitem[\protect\citename{Chiang \& Murray} 1996]{chi96}
    Chiang J., Murray N.,
    1996, ApJ, 466, 704.
   
  \bibitem[\protect\citename{Czerny \& Hryniewicz} 2011]{cze11}
    Czerny, B., Hryniewicz, K.
    2011, A\&A, 525L, 8.   
     
  \bibitem[\protect\citename{Denney et al.} 2006]{den06}
    Denney K. D., et al.
    2009, ApJ, 704, L80.
    
  \bibitem[\protect\citename{Done \& Krolik} 1996]{don06}
    Done C., \& Krolik J. H. 
    1996, ApJ, 463, 144.
  
  \bibitem[\protect\citename{Elvis} 2000]{el00}
    Elvis, M.
    2000, ApJ, 545, 63.
  
  \bibitem[\protect\citename{Elitzur} 2008]{elit08}
    Elitzur, M. 
    2008, NewAR, 52, 274.
 
  \bibitem[\protect\citename{Gaskell} 2009]{gas09}
    Gaskell M.,
    2009, New Astr. Rev., 53, 140.
   
  \bibitem[\protect\citename{Kaspi et al.} 2000]{kas00}
    Kaspi et al.,
    2000, ApJ, 533, 631.
    
  \bibitem[\protect\citename{Marconi et al.} 2008]{mar08}
    Marconi A. et al.,
    2008, ApJ, 678, 693.
  
  \bibitem[\protect\citename{Peterson et al.} 2004]{pet04}
    Peterson et al.,
    2004, ApJ, 613, 682.
    
  \bibitem[\protect\citename{Shakura \& Sunyaev} 1973]{sug73}
    Shakura, N. I.; Sunyaev, R. A.
    1973, A\&A, 24, 337.
   
  \bibitem[\protect\citename{Suganuma et al.} 2006]{sug06}
    Suganuma M., et al.
    2006, ApJ, 639, 46.
    
  \bibitem[\protect\citename{Watson et al.} 2011]{wat11}
    Watson, D. et al.
    2011, ApJ, 740, L49.

\end{thebibliography}
\end{document}